\documentclass[11pt]{article}
\usepackage{mathrsfs}
\usepackage{graphicx}
\usepackage{amsfonts}
\begin{document}
\vspace*{-.6in} \thispagestyle{empty}
\begin{flushright}
hep-th/0612144
\end{flushright}
\baselineskip = 18pt

\vspace{.5in} {\Large
\begin{center}
\textbf{Thermodynamics of Apparent Horizon in Brane World Scenario}
\end{center}}
\vspace*{1cm}

\begin{center}
Rong-Gen Cai,$^{a,}$\footnote{Email address: cairg@itp.ac.cn} and
Li-Ming Cao$^{a,b,}$\footnote{Email address: caolm@itp.ac.cn}
\end{center}

\begin{center}
\emph{$^{a}$ Institute of Theoretical Physics, Chinese Academy
of Sciences, \\ P.O. Box 2735, Beijing 100080, China \\
$^{b}$ Graduate School of the Chinese Academy of Sciences, Beijing
100039, China }

\vspace{.5in}


\end{center}

\vspace{.2in}

\begin{center}
\underline{ABSTRACT}
\end{center}
\begin{quotation}
\noindent In this paper we discuss thermodynamics of apparent
horizon of an $n$-dimensional Friedmann-Robertson-Walker (FRW)
universe embedded in an $(n+1)$-dimensional AdS spacetime. By
using the method of unified first law, we give the explicit
entropy expression of the apparent horizon of the FRW universe. In
the large horizon radius limit, this entropy reduces to the
$n$-dimensional area formula, while in the small horizon radius
limit, it obeys the $(n+1)$-dimensional area formula. We also
discuss the corresponding bulk geometry and study the apparent
horizon extended into the bulk. We calculate the entropy of this
apparent horizon by using the area formula of the
$(n+1)$-dimensional bulk. It turns out that both methods give the
same result for the apparent horizon entropy. In addition, we show
that the Friedmann equation on the brane can be rewritten to a
form of the first law, $dE=TdS +WdV$, at the apparent horizon.

\end{quotation}

\vfil \centerline{}

\newpage

\addtocontents{toc}{\protect\setcounter{tocdepth}{2}}
\pagenumbering{arabic}

\vspace{.5in}
\section{Introduction}
Thermodynamics of black hole has been studied for a long time.
However, most discussions of black hole thermodynamics have been
focused on the stationary case.  For dynamical (i.e.,
non-stationary) spherically symmetric black holes, Hayward has
proposed a method to deal with thermodynamics associated with
trapping horizon of a dynamical black hole in $4$-dimensional
Einstein theory~\cite{Hayward, Hayward1, Hayward2, Hayward3}. In
this method, for spherical symmetric space-times, $ds^2
=h_{\alpha\beta}dx^{\alpha} dx^{\beta} + r^2 d\Omega_2^2$,
Einstein equations can be rewritten in a form called ``unified
first law"
\begin{equation}
\label{eq1}
 dE=A\Psi+WdV\, ,
\end{equation}
where $E$ is the so-called Misner-Sharp energy~\cite{MS}, defined
by $E=\frac{r}{2G}(1-h^{\alpha\beta}\partial_{\alpha}r
\partial_{\beta}r)$; $A=4\pi r^2$ is the sphere area with radius $r$ and
$V=4\pi r^3/3$ is the volume; and the work density $W \equiv
-\frac{1}{2}T^{\alpha\beta}h_{\alpha\beta}$ and the energy supply
vector $\Psi_{\alpha} \equiv T_{\alpha}^{\ \beta}\partial_{\beta}r
+W
\partial_{\alpha}r$ with $T_{\alpha\beta}$ being the
energy-momentum tensor of matter in the spacetime.
 Projecting this unified first law along a trapping
horizon, one gets the first law of thermodynamics for dynamical
black hole
\begin{equation}
\label{eq2}
 \langle dE,\xi\rangle=\frac{\kappa}{8\pi G}\langle
dA,\xi\rangle+W\langle dV,\xi\rangle\, ,
\end{equation}
where  $\kappa$ is surface gravity, defined by $\kappa
=\frac{1}{2\sqrt{-h}}\partial_{\alpha}(\sqrt{-h}h^{\alpha\beta}
\partial_{\beta}r)$,
on the apparent horizon, $\xi$ is a projecting vector. In a recent
paper~\cite{caicao}, we have applied this theory to study the
thermodynamics of apparent horizon of a FRW universe in higher
dimensional Einstein gravity and some non-Einstein theories, such
as Lovelock gravity and scalar-tensor gravity by rewriting gravity
field equations to standard Einstein equations with a total
energy-momentum tensor.  The total energy momentum tensor consists
of two parts: one is just the ordinary matter energy-momentum
tensor; and the other is an effective one coming from the
contribution of the higher derivative terms (for example, in
Lovelock gravity) and scalar field (for example, in scalar-tensor
gravity theory). The total energy-momentum tensor will enter into
the energy-supply $\Psi$ and work term $W$ in (\ref{eq1}).
Therefore, the work density and energy-supply vector in
(\ref{eq1}) also can be decomposed into ordinary matter part and
effective part: $\Psi=\stackrel{m}{\Psi}+\stackrel{e}{\Psi}$,
$W=\stackrel{m}{W}+\stackrel{e}{W}$. That is, $
\stackrel{m}{\Psi}$ and $\stackrel{m}{W}$ are the energy-supply
vector and work density from the ordinary matter contribution,
while $\stackrel{e}{\Psi}$ and $\stackrel{e}{W}$ comes from the
effective energy-momentum tensor. The matter part of energy-supply
is the energy flux defined by pure ordinary matter, so its
integration on the sphere (after projecting along the apparent
horizon) naturally defines the heat flow $\delta Q $ in the
Clausius relation $\delta Q=T dS$.  On the other hand, the unified
first law tells us
\begin{equation}
\delta Q=\langle A\stackrel{m}{\Psi} ,\xi\rangle=\frac{\kappa}{8\pi
G}\langle dA,\xi\rangle-\langle A\stackrel{e}{\Psi} ,\xi\rangle\, .
\end{equation}
The right hand side of the above equation should be written in the
form of $T\delta S$ by considering spacetime horizon with
$T=\kappa/2\pi$ as an  equilibrium thermodynamic system. Thus we
find a method to get the entropy $S$ of the apparent horizon
because the right hand side of the above equation is easy to
calculate. What one needs to do is to put the right hand side of
the above equation into a form of  total differential projecting
along $\xi$. This total differential gives the variation of the
horizon entropy $S$. By using this method we have indeed given the
 entropy expression not only in the Einstein gravity, but also in the
 Lovelock gravity. The resulting expression of the apparent horizon
 entropy is the same as the one for black hole horizon in each theory.
For related discussions on the thermodynamics of apparent horizon
in the FRW universe, see \cite{a11}-\cite{acai2}. In the setup of
static, spherically symmetric black hole spacetimes,  there are
also some discussions on the relation between the field equations
at the black hole horizon and the first law of
thermodynamics~\cite{Pad,pad1,acai2}. For the Rindler causal
horizon, related discussions see \cite{jac1,jac2}.

 It is interesting to apply this method developed in \cite{caicao} to
study the entropy of the apparent horizon of a FRW universe in the
brane world scenario. This is partially because the gravity on the
brane is not the Einstein theory, the well-known area formula for
black hole horizon entropy must not hold in this case, and
partially because exact analytic black hole solutions on the brane
have not been found so far, and then it is not known how the
horizon entropy of black hole on the brane is determined by the
horizon geometry. On the other hand,  the exact Friedmann
equations of the FRW universe on the brane have been derived for
the RSII model some years ago~\cite{Langlois}. Therefore with the
entropy expression of the apparent horizon by using the Friedmann
equations,  our method can give some clues to study the
thermodynamics of the black holes on the brane. In this paper,
 we are indeed able to give the explicit entropy expression
of the apparent horizon of FRW universe in the RSII  brane world
scenario.

Brane world scenario, based on the assumption that our universe is
a $3$-brane embedded in a higher dimensional bulk space-time, has
been intensively studied over past
years~\cite{ADD,Randall1,Randall2}. In the scenario, the standard
model fields are confined on the brane, while gravity can
propagates in the whole spacetime. The effective gravity on the
brane is different from the standard Einstein gravity due to the
existence of extra dimension.  In this paper we focus on the
so-called RSII model~\cite{Randall2}. The effective equations of
motion on the $3$-brane living in $5$-dimensional bulk with
$\mathbb{Z}_2$ symmetry have been given in~\cite{Shiromizu}
\begin{eqnarray}
\label{eeq4} {}^{(4)}G_{\mu\nu}=-\Lambda_4 q_{\mu\nu} + 8 \pi
G_4\tau_{\mu\nu}+\kappa_{5}^4\,\pi_{\mu\nu} -E_{\mu\nu}\,,
\end{eqnarray}
where
\begin{equation}
\label{en4} G_4=\frac{1}{48\pi}\lambda\kappa_{5}^4\,,
\end{equation}
\begin{equation}
\label{ecc4} \Lambda_4=\frac{1}{2}\kappa_{5}^2 \left(\Lambda_{5}
+\frac{1}{6}\kappa_{5}^2\,\lambda^2\right)\,,
\end{equation}
\begin{equation}
\label{eem4} \pi_{\mu\nu}= -\frac{1}{4}
\tau_{\mu\alpha}\tau_\nu^{~\alpha} +\frac{1}{12}\tau\tau_{\mu\nu}
+\frac{1}{8}q_{\mu\nu}\tau_{\alpha\beta}\tau^{\alpha\beta}-\frac{1}{24}
q_{\mu\nu}\tau^2\,, \label{pidef4}
\end{equation}
and $E_{\mu\nu}$ is the electric part of the $5$-dimensional Weyl
tensor. Here $\lambda$ and $\tau_{\mu\nu}$ are the vacuum energy
and energy-momentum tensor of matter on the brane, while
$\kappa_5$ and $\Lambda_5$ are $5$-dimensional gravity coupling
constant and cosmological constant, respectively. $\Lambda_4$ is
the effective cosmological constant on the brane. Therefore it can
be seen from the right hand side of (\ref{eeq4}) that except for
the energy-momentum tensor of matter $\tau_{\mu\nu}$, there exist
two additional terms $\pi_{\mu\nu}$ and $E_{\mu\nu}$. This implies
that these two terms can be regarded as effective energy-momentum
tensors, which can not enter the definition of $\delta Q$ in the
Clausius relation as mentioned above. If the bulk is a pure AdS
space-time, then $E_{\mu\nu}$ vanishes, and $\pi_{\mu\nu}$ is the
only effective energy-momentum tensor. Thus, we can use our method
to obtain the entropy expression of apparent horizon in the FRW
universe on the brane. We expect that this entropy expression
should reveal some
 properties of the horizon entropy of brane world black hole.
Indeed our result is consistent with the one in
reference~\cite{Emparan}, where the authors give a horizon entropy
of an $n$-dimensional black hole on a brane, which is embedded in an
$(n+1)$-dimensional bulk, in the limit of large horizon radius.

The apparent horizon of FRW universe on the brane will extend into
the  bulk (AdS space-time) with a finite distance. The total area
of this apparent horizon which extends into the bulk can be
directly calculated from the bulk geometry. Since the gravity in
the bulk is the Einstein's general relativity, the well-known area
formula of horizon entropy holds. Therefore, from the higher
dimensional area formula of entropy we can get an entropy of this
apparent horizon. We find that the horizon entropy, according to
the area formula in the bulk, is  completely the same as the one
obtained by using the unified first law on the brane.

The paper is organized as follows. In Sec. 2, we give the
effective equations of motion on the $(n-1)$-brane embedded in an
$(n+1)$-dimensional bulk with $\mathbb{Z}_2$ symmetry by
generalizing the result of \cite{Shiromizu}. In Sec. 3, we
consider a FRW universe on the brane, and calculate the entropy of
 apparent horizon by use of the unified first law. In Sec. 4, we study
the bulk geometry and the apparent horizon extended into the bulk.
By calculating the area of this apparent horizon extended into the
bulk, we get the same entropy from the $(n+1)$-dimensional area
formula as the one obtained in Sec.~3. We end this paper with
conclusion in Sec.~5.

\section{Effective equations of motion on the brane}

In the brane world scenario, the $n$-dimensional world is
described by an $(n-1)$-brane $(\mathscr{M},q_{\mu\nu})$ in an
$(n+1)$-dimensional spacetime $(\mathscr{V},g_{\mu\nu})$. We
denote the vector unit normal to $\mathscr{M}$ by $n^\alpha$ and
the induced metric on $\mathscr{M}$ by $q_{\mu\nu} = g_{\mu\nu} -
n_{\mu}n_{\nu}$ (we use the notations in~\cite{Shiromizu}). Then
from the Gauss equation
\begin{equation}
{}^{(n)}R^\alpha_{~\beta\gamma\delta} =
{}^{(n+1)}R^\mu_{~\nu\rho\sigma} q^{~\alpha}_\mu q_\beta^{~\nu}
q_\gamma^{~\rho}
 q_\delta^{~\sigma} + K^\alpha_{~\gamma}K_{\beta\delta}
-K^\alpha_{~\delta}K_{\beta\gamma}\,, \label{Gauss}
\end{equation}
and $(n+1)$-dimensional Einstein equations
\begin{equation}
{}^{(n+1)}R_{\alpha\beta} -\frac{1}{2}g_{\alpha\beta}{}^{(n+1)}R
=\kappa_{n+1}^2\, T_{\alpha\beta}\, , \label{n+1dEinstein}
\end{equation}
where  $T_{\alpha\beta}$ is the $(n+1)$-dimensional
energy-momentum tensor in the bulk, together with  the relation of
Weyl tensor and Riemann tensor, we obtain the $n$-dimensional
equations on the brane
\begin{eqnarray}
{}^{(n)}G_{\mu\nu} &=& \frac{n-2
}{n-1}\kappa_{n+1}^2\left(T_{\rho\sigma} q^{~\rho}_{\mu}
q^{~\sigma}_{\nu} +\left(T_{\rho\sigma}n^\rho
n^\sigma-\frac{1}{n}T^\rho_{~\rho}\right)
 q_{\mu\nu} \right)\nonumber \\
&&+ KK_{\mu\nu} -K^{~\sigma}_{\mu}K_{\nu\sigma} -{1 \over
2}q_{\mu\nu}
  \left(K^2-K^{\alpha\beta}K_{\alpha\beta}\right) - E_{\mu\nu},
\label{ndEinstein}
\end{eqnarray}
where
\begin{equation}
E_{\mu\nu} \equiv {}^{(n+1)}C^\alpha_{~\beta\rho\sigma}n_\alpha
n^\rho q_\mu^{~\beta} q_\nu^{~\sigma} . \label{Edef}
\end{equation}
is the electric part of bulk Weyl tensor. We choose a local Gauss
normal coordinate $\chi$ such that the hypersurface $\chi=0$
coincides with the brane world and $n_\mu dx^\mu=d\chi$. Assume that
the bulk has only a cosmological constant $\Lambda_{n+1}$, the
$(n+1)$-dimensional energy-momentum tensor then has the form
\begin{eqnarray}
T_{\mu\nu}=-\Lambda_{n+1} g_{\mu\nu}+S_{\mu\nu}\delta (\chi),
\label{eq:bulk}
\end{eqnarray}
where
\begin{eqnarray}
S_{\mu\nu}=-\lambda q_{\mu\nu}+\tau_{\mu\nu}\,, \label{eq:matter}
\end{eqnarray}
with $\tau_{\mu\nu}n^\nu=0$, $\lambda$ and $\tau_{\mu\nu}$ are the
brane tension and the energy-momentum tensor of matter on the
$(n-1)$-brane. The singular behavior in the energy-momentum tensor
can be attributed to extrinsic curvature from the so-called Israel's
junction condition
\begin{equation}
\left[q_{\mu\nu}\right] =0\,,
\end{equation}
\begin{equation}
\left[K_{\mu\nu}\right]= -\kappa_{n+1}^2
\Bigl(S_{\mu\nu}-\frac{1}{n-1}q_{\mu\nu}S\Bigr),
\end{equation}
where $[X]:=\lim_{\chi \to +0}X - \lim_{\chi \to -0}X=X^+-X^-$.

After imposing the $\mathbb{Z}_2$-symmetry on the bulk spacetime,
the extrinsic curvature of the brane can be expressed in terms of
the energy momentum tensor on the brane
\begin{eqnarray}
K_{\mu\nu}^+=-K^-_{\mu\nu}=-\frac{1}{2}\kappa_{n+1}^2
\Bigl(S_{\mu\nu}-\frac{1}{n-1}q_{\mu\nu}S\Bigr)\,. \label{eqzsym}
\end{eqnarray}
Substituting this equation (the $\pm$ is not relevant because of
the $\mathbb{Z}_2$ symmetry) into Eq.~(\ref{ndEinstein}), we
obtain the gravitational field equations on the $(n-1)$-brane
\begin{eqnarray}
{}^{(n)}G_{\mu\nu}=-\Lambda_n q_{\mu\nu} + 8 \pi
G_n\tau_{\mu\nu}+\kappa_{n+1}^4\,\pi_{\mu\nu} -E_{\mu\nu}\,,
\label{eq:effective}
\end{eqnarray}
where
\begin{equation}
G_n=\frac{n-2}{32\pi(n-1)}\lambda\kappa_{n+1}^4\,, \label{GNdef}
\end{equation}
\begin{equation}
\Lambda_n=\kappa_{n+1}^2 \left(\frac{n-2}{n}\Lambda_{n+1}
+\frac{n-2}{8(n-1)}\kappa_{n+1}^2\,\lambda^2\right)\,,
\label{Lamda4}
\end{equation}
\begin{equation}
\pi_{\mu\nu}= -\frac{1}{4} \tau_{\mu\alpha}\tau_\nu^{~\alpha}
+\frac{1}{4(n-1)}\tau\tau_{\mu\nu}
+\frac{1}{8}q_{\mu\nu}\tau_{\alpha\beta}\tau^{\alpha\beta}-\frac{1}{8(n-1)}
q_{\mu\nu}\tau^2\,, \label{pidef}
\end{equation}
and $E_{\mu\nu}$ is the electric part of the $(n+1)$-dimensional
Weyl tensor defined in Eq.~({\ref{Edef}). The above equations
reduce to the ones in~\cite{Shiromizu} as $n=4$, as expected.
These equations (\ref{eq:effective}) are not closed  since the
bulk geometry can not be determined by using only these equations
without solving the Einstein equations in the bulk. The standard
Einstein equations can be recovered by taking the limit
$\kappa_{n+1}\to 0$ while keeping $G_n$ finite. Hereafter, we
specialize to the RS II model with vanishing cosmological constant
on the brane, which gives
\begin{equation}
\Lambda_{n+1}=-\frac{n(n-1)}{2\kappa_{n+1}^2\ell^2}\, ,\quad
\lambda=\frac{2(n-1)}{ \kappa_{n+1}^2\ell}\, .
\end{equation}
 This also implies
\begin{equation}
\label{n+1ton}
 \frac{G_{n+1}}{G_{n}}=\frac{ \kappa_{n+1}^2}{8\pi
 G_{n}}=\frac{2\ell}{n-2}\, .
\end{equation}

Now we consider a FRW universe with perfect fluid confined on the
$(n-1)$-brane.  The perfect fluid has the  energy-momentum tensor
\begin{equation}
\tau_{\mu\nu}=(\rho+p)t_{\mu}t_{\nu}+pq_{\mu\nu}\, ,
\end{equation}
where $t_{\mu}$ satisfies $t^{\mu}t_{\mu}=-1$. By using the
expression of $\pi_{\mu\nu}$, we find
\begin{equation}
\pi_{\mu\nu}=\frac{n-2}{8(n-1)}\rho\left[2(\rho+p)
t_{\mu}t_{\nu}+(\rho+2p)q_{\mu\nu}\right]\, .
\end{equation}
The equations of motion on the brane then become
\begin{eqnarray}
\label{e25}
 {}^{(n)}G_{\mu\nu}&=& 8 \pi
G_n\left[\rho+p+\frac{n-2}{n-1}\frac{\kappa_{n+1}^4}{32 \pi
G_n}\rho(\rho+p)\right]t_{\mu}t_{\nu}\nonumber \\
&&+ 8 \pi G_n\left[p+\frac{n-2}{n-1}\frac{\kappa_{n+1}^4}{64 \pi
G_n}\rho(\rho+2p)\right]q_{\mu\nu}-E_{\mu\nu} \, .
\end{eqnarray}
For a locally conformal flat bulk spacetime (for example, a pure
AdS spacetime), the bulk Weyl tensor  vanishes, so we can omit the
$E_{\mu\nu}$ term in that case. Choosing $q_{\mu\nu}$ to be
$n$-dimensional FRW metric (FRW universe can indeed be embedded in
a pure AdS space-time), i.e.,
\begin{eqnarray}
ds^2&=&q_{\mu\nu}dx^{\mu}dx^{\nu}=-dt^2+\frac{a(t)^2}{1-kr^2}dr^2+a(t)^2r^2d\Omega_{n-2}^2
\nonumber\\
&=&\tilde{h}_{ab}dx^a dx^b+\tilde{r}^2d\Omega_{n-2}^2\, ,
\end{eqnarray}
where $\tilde{r}=a(t)r$, $x^0=t, x^1=r$,  from (\ref{e25}) we obtain
the Friedmann equations on the brane without dark radiation term
($E_{\mu\nu}=0$)
\begin{equation}
\label{e27}
 (n-1)(n-2)\left(H^2+\frac{k}{a^2}\right)=16\pi G_n\left(
\rho+\frac{n-2}{n-1}\frac{\kappa_{n+1}^4}{64 \pi G_n}\rho^2\right)\,
,
\end{equation}
\begin{equation}
\label{e28}
 -(n-2)\left(\dot{H}-\frac{k}{a^2}\right) =8\pi G_n
\left(\rho+p+\frac{n-2}{n-1}\frac{\kappa_{n+1}^4}{32 \pi
G_n}\rho(\rho+p) \right)\, .
\end{equation}
The matter density $\rho$ on the brane satisfies the continuity
equation
\begin{equation}
\label{e29}
 \dot{\rho}+(n-1)H(\rho+p)=0\, .
\end{equation}
This equation and the Friedmann equations will be used in the next
sections.

\section{Unified first law and thermodynamics of apparent horizon}

In this section we will give the explicit entropy expression of
apparent horizon in a FRW universe on  the brane by using  the
method developed in \cite{caicao}. Introducing the effective
energy-momentum tensor
\begin{equation}
\stackrel{e}{\tau}_{\mu\nu}=\frac{\kappa_{n+1}^4}{8 \pi
G_n}\pi_{\mu\nu}\, .
\end{equation}
We can rewrite the gravitational field equations on the brane as
\begin{equation}
\label{e31}
 {}^{(n)}G_{\mu\nu}= 8 \pi
G_n\left(\tau_{\mu\nu}+\stackrel{e}{\tau}_{\mu\nu}\right)\, ,
\end{equation}
which is in the form of the standard Einstein field equations.
Therefore the unified first law (\ref{eq1}) is applicable to the
equation (\ref{e31}).  It is easy to find
\begin{equation}
\stackrel{e}{\tau}_{t}{}^{t}=-\frac{n-2}{n-1}\frac{\kappa_{n+1}^4}{64\pi
G_{n}}\rho^2\, ,\quad\quad
\stackrel{e}{\tau}_{r}{}^{r}=\frac{n-2}{n-1}\frac{\kappa_{n+1}^4}{64\pi
G_{n}} \rho(\rho+2p)\, .
\end{equation}
The work density term has the form
\begin{equation}
W=\stackrel{m}{W}+\stackrel{e}{W},
\end{equation}
with
\begin{equation}
\stackrel{m}{W}=\frac{1}{2}(\rho-p)\, ,\quad\quad
\stackrel{e}{W}=-\frac{n-2}{n-1}\frac{\kappa_{n+1}^4}{64\pi G_{n}}
\rho p\, .
\end{equation}
Note that quantities with over $m$ are calculated through the
matter energy-momentum tensor $\tau_{\mu\nu}$, while quantities
with over $e$ are calculated through the effective energy-momentum
tensor $\stackrel{e}{\tau}_{\mu\nu}$. Namely,
$\stackrel{m}{W}=-\frac{1}{2}\tau^{\mu\nu}\tilde{h}_{\mu\nu}$, and
$ \stackrel{e}{W}=
-\frac{1}{2}\stackrel{e}{\tau}_{\mu\nu}\tilde{h}^{\mu\nu}$.
Similarly, the energy supply vector can be decomposed as
\begin{equation}
\Psi=\stackrel{m}{\Psi}+\stackrel{e}{\Psi},
\end{equation}
with
\begin{equation}
\stackrel{m}{\Psi}=-\frac{1}{2}(\rho+p)H\tilde{r}dt+\frac{1}{2}(\rho+p)adr\,
,
\end{equation}
\begin{equation}
\stackrel{e}{\Psi}=-\frac{n-2}{n-1}\frac{\kappa_{n+1}^4}{64\pi
G_{n}}\rho(\rho+p)H\tilde{r}dt+\frac{n-2}{n-1}\frac{\kappa_{n+1}^4}{64\pi
G_{n}}\rho(\rho+p)adr\,,
\end{equation}
where they are defined by $\stackrel{m}{\Psi}=\tau \partial \tilde
r + \stackrel{m}W \partial \tilde r $ and $\stackrel
{e}{\Psi}=\stackrel{e}{\tau} \partial \tilde r
+\stackrel{e}{W}\partial \tilde r$, respectively.  Using these
quantities and the Misner-Sharp energy in $n$ dimensions inside
the apparent horizon~\cite{BR, caicao}
\begin{equation}
E=\frac{1}{16\pi G_n}(n-2)\Omega_{n-2}\tilde{r}_A^{n-3}\, ,
\end{equation}
and $A_{n-2}=\Omega_{n-2}\tilde r_A^{n-2}$ and $V_{n-2}=A_{n-2}
\tilde{r}_A/(n-1)$ are the area and volume of the $(n-2)$- sphere
with the apparent horizon radius $\tilde{r}_A=1/\sqrt{H^2+k/a^2}$
of the FRW universe, respectively, we can put the $(00)$ component
of equations of motion (\ref{e31})  into the form of the unified
first law
\begin{equation}
dE=A_{n-2}\Psi+WdV_{n-2}\, .
\end{equation}
After projecting along a vector
$\xi=\partial_t-(1-2\epsilon)Hr\partial_r$ with
$\epsilon=\dot{\tilde{r}}_A/2H\tilde{r}_A$, we get the first law of
thermodynamics of the apparent horizon~\cite{caicao},
\begin{equation}
\langle dE, \xi\rangle=\frac{\kappa}{8\pi G_n}\langle dA_{n-2},
\xi\rangle+\langle WdV_{n-2}, \xi\rangle\,,
\end{equation}
where $\kappa=-(1-\dot {\tilde r}_A/(2H\tilde r_A))/\tilde r_A$ is
the surface gravity of the apparent horizon. The pure matter
energy-supply $A_{n-2}\stackrel{m}{\Psi}$ (after projecting along
the apparent horizon) gives the heat flow $\delta Q$ in the Clausius
relation $\delta Q=T d S$. By using the unified first law on the
apparent horizon, we have
\begin{eqnarray}
\delta Q& \equiv &\langle
A_{n-2}\stackrel{m}{\Psi},\xi\rangle=\frac{\kappa}{8\pi G_n}\langle
dA_{n-2},\xi \rangle-\langle
A_{n-2}\stackrel{e}{\Psi},\xi\rangle, \nonumber \\
&=&\frac{\kappa}{8\pi G_n}\langle dA_{n-2},\xi
\rangle+\frac{n-2}{n-1}\frac{\kappa_{n+1}^4}{32\pi
G_{n}}(1-\epsilon)\rho(\rho+p)A_{n-2}H\tilde{r}_A\, .
\end{eqnarray}
From the  Friedmann equations given in the previous section, one
finds
\begin{equation}
\rho(\rho+p)=\frac{8(n-1)\epsilon}{\kappa_{n+1}^4}
\left(\frac{1}{\tilde{r}_A^2}-\frac{1}{\sqrt{\tilde{r}_A^4+\ell^2\tilde{r}_A^2}}\right)\,
.
\end{equation}
Therefore, we arrive at
\begin{eqnarray}
\delta Q &=&-\frac{\epsilon(1-\epsilon)}{4\pi
G_n}(n-2)A_{n-2}H\tilde{r}_A\left(\frac{1}{\sqrt{\tilde{r}_A^4+\ell^2\tilde{r}_A^2}}\right)\nonumber\\
&=&T\langle \frac{ n-2
}{4G_{n}}\frac{\Omega_{n-2}\tilde{r}_A^{n-2}}{\sqrt{\tilde{r}_A^2+\ell^2}}d\tilde{r}_A,
\xi \rangle=T\langle dS,\xi \rangle=TdS\, ,
\end{eqnarray}
where
\begin{equation}
\label{e44}
S=\frac{(n-2)\Omega_{n-2}}{4G_n}{\displaystyle\int^{\tilde
r_A}_0\frac{\tilde{r}_A^{n-2}
}{\sqrt{\tilde{r}_A^2+\ell^2}}d\tilde{r}_A}\,.
\end{equation}
Integrating (\ref{e44}), we obtain the entropy expression associated
with the apparent horizon
\begin{equation}
\label{entropy} \label{e45}
S=\frac{A_{n-2}}{4G_n}\cdot\left\{\frac{n-2}{n-1}\left(\frac{\tilde{r}_A}{\ell}\right)
{}_2F_1\left[\frac{n-1}{2},\frac{1}{2},\frac{n+1}{2},
-\left(\frac{\tilde{r}_A}{\ell}\right)^2\right]\right\}\, ,
\end{equation}
where ${}_2F_1[\alpha,\beta,\gamma,z]$ is Gaussian hypergeometric
function. When $n=4$, we have
\begin{equation}
\label{e46}
S=\frac{A_2}{4G_4}\left[\sqrt{1+\frac{\ell^2}{\tilde{r}_A^2}}
-\frac{\ell^2}{\tilde{r}_A^2}\mathrm{arcsinh}\left(\frac{\tilde{r}_A}{\ell}\right)\right]\,
.
\end{equation}
Here some remarks are in order.

 (i). Taking the limit $\kappa_{n+1}\rightarrow 0$
while keeping $G_{n}$ finite, we have $\ell\rightarrow 0$. In this
limit, the $n$-dimensional Einstein gravity is recovered on the
brane. This limit can also be understood as the one of the large
apparent horizon radius, namely, $\tilde r_A \gg \ell$. In this
limit, we see from (\ref{e45}) that the entropy expression of the
apparent horizon reduces to the well-known area formula of horizon
entropy in $n$ dimensions.  This is an expected result since the
gravity is an Einstein one in this limit, where the area formula
holds.  The entropy can also be expressed by
\begin{equation}
S=\frac{A_{n-2}}{4G_n}=\frac{A_{n-2}}{4G_{n+1}}\left(\frac{2\ell}{n-2}\right)\,
.
\end{equation}
This indicates that in the limit of large horizon radius, the
entropy is proportional to the area of the ``cylinder" with length
$2\ell/(n-2)$ if we look at it from the viewpoint of the bulk. This
coincides with the argument in~\cite{Emparan}.

 (ii). For small $\tilde{r}_A \ll \ell $, we can expand the
hypergeometric function, and reach
\begin{equation}
S=\frac{A_{n-2}}{4G_n}\cdot\left\{\frac{n-2}{n-1}\frac{\tilde{r}_A}{\ell}+\mathcal
{O}(\tilde{r}_A^2)\right\}\, .
\end{equation}
Using the relation (\ref{n+1ton}), we have, up to the first order,
\begin{equation}
S=\frac{A_{n-2}}{4G_n}\cdot\frac{n-2}{n-1}\frac{\tilde{r}_A}{\ell}
=\frac{1}{4G_{n+1}}\frac{2A_{n-2}\tilde{r}_A}{n-1}\, .
\end{equation}
Note that the volume of the $(n-2)$-dimensional sphere with horizon
radius is $V_{n-2}=A_{n-2}\tilde{r}_A/(n-1) $,  we find the entropy
in the small apparent horizon limit
\begin{equation}
S =\frac{2V_{n-2}}{4G_{n+1}}\, .
\end{equation}
Namely, this entropy produces the $(n+1)$-dimensional properties
of theory, and obeys the area formula of horizon in ($n+1$)
dimensions. The factor $2$ is due to the $\mathbb{Z}_2$ symmetry
of the bulk. Therefore, in the small apparent horizon limit, the
$(n+1)$-dimensional effect will be remarkable. The $V_{n-2}$ is
just the volume of the ``disk" inside the sphere which has area
$A_{n-2}$. The areas of these two ``disks" are negligible under
the large horizon radius limit~\cite{Emparan}, while they are
important in the small radius limit.

 (iii). The unified first law can be rewritten to be
\begin{equation}
dE-A_{n-2}\stackrel{e}{\Psi}-\stackrel{e}{W}dV_{n-2}=A_{n-2}\stackrel{m}
{\Psi}+\stackrel{m}{W}dV_{n-2}\,
.
\end{equation}
By using the first, second Friedmann equations, (\ref{e27}) and
(\ref{e28}), and  the continuity equation (\ref{e29}), we find
\begin{equation}
dE-A_{n-2}\stackrel{e}{\Psi}-\stackrel{e}{W}dV_{n-2}=d(\rho
V_{n-2})\, .
\end{equation}
Here $\rho V_{n-2}$  is nothing, but the total energy of matter
inside the apparent horizon.  Thus we can rewrite the unified first
law as
\begin{equation}
\label{uflpurematter}
d\stackrel{m}{E}=A_{n-2}\stackrel{m}{\Psi}+\stackrel{m}{W}dV_{n-2},
\end{equation}
where $\stackrel{m}{E}=\rho V_{n-2}$. After projecting along the
horizon using $\xi$, equation (\ref{uflpurematter}) gives us with
\begin{equation}
\langle d\stackrel{m}{E},\xi\rangle=\langle TdS,
\xi\rangle+\langle\stackrel{m}{W}dV_{n-2},\xi\rangle.
\end{equation}
This relation is nothing but the first law of thermodynamics
associated with the apparent horizon~\cite{acai1},
$dE=TdS+WdV_{n-2}$, with identifying the inner energy $E$ to be
$\rho V_{n-2}$, temperature $T$ to $\kappa/2\pi$, entropy $S$ to the
form given in~(\ref{entropy}), and the work density $W=(\rho-p)/2$.

 (iv). If the bulk Weyl tensor does not vanish, the thing
becomes complicated. Without the knowledge of bulk geometry, we can
not obtain an entropy expression of apparent horizon in terms of the
horizon radius.

\section{Bulk geometry and apparent horizon extended into bulk }

From the global point of view, the apparent horizon on the brane
will extend into the bulk. And then the entropy of the apparent
horizon can be determined by the area formula in the bulk (here we
have assumed that the gravity is the Einstein one in the bulk). In
this section, we will directly calculate the area of the apparent
horizon which extends into the bulk, and give the entropy
expression of apparent horizon from the bulk geometry.  For this
aim, we have to first find out the bulk geometry. Assume that the
metric in the bulk has the form
\begin{equation}
\label{bulkmetric}
d\tilde{s}^2=-N^2(t,z)dt^2+\frac{A^2(t,z)}{1-kr^2}dr^2+A^2(t,z)r^2d\Omega_{n-2}^2+dz^2\,
,
\end{equation}
where $A$ and $N$ satisfy (the brane is supposed to locate at
$z=0$
 with $\mathbb{Z}_2$ symmetry)
\begin{equation}
N(t,0)=1\, , \quad A(t,0)=a(t)\, , \quad N(t,z)=N(t,-z)\,
,A(t,z)=A(t,-z)\, ,
\end{equation}
Then we have the nonvanishing components of Einstein tensor
\begin{equation}
{}^{(n+1)}G_{0z}=(n-1)\left[\frac{N'}{N}\frac{\dot{A}}{A}-\frac{\dot{A}'}{A}\right]\,
,
\end{equation}
\begin{equation}
{}^{(n+1)}G_{00}=\frac{1}{2}(n-1)(n-2)\left[\frac{\dot{A}^2}{A^2}-N^2\left(\frac{{A'}^2}{A^2}
+\frac{2}{n-2}\frac{A''}{A}\right)+k\frac{N^2}{A^2}\right]\, ,
\end{equation}
\begin{equation}
{}^{(n+1)}G_{zz}=\frac{1}{2}(n-1)(n-2)\left[\frac{{A'}^2}{A^2}+
\frac{2}{n-2}\frac{A'}{A}\frac{N'}{N}-\frac{1}{N^2}\left(\frac{\dot{A}^2}{A^2}
-\frac{2}{n-2}\frac{\dot{A}}{A}\frac{\dot{N}}{N}
+\frac{2}{n-2}\frac{\ddot{A}}{A}\right)-\frac{k}{A^2}\right]\, ,
\end{equation}
\begin{eqnarray}
{}^{(n+1)}G_{ij}&=&\frac{1}{2}(n-2)(n-3)A^2\gamma_{ij}\Bigg{[}
\frac{{A'}^2}{A^2}+\frac{2}{n-3}\frac{A'}{A}\frac{N'}{N}+\frac{2}{n-3}\frac{A''}{A}
+\frac{2}{(n-2)(n-3)}\frac{N''}{N}\nonumber
\\
&+&\frac{1}{N^2}\left(-\frac{\dot{A}^2}{A^2}+\frac{2}{n-3}\frac{\dot{N}}{N}\frac{\dot{A}}{A}
-\frac{2}{n-3}\frac{\ddot{A}}{A}\right)-\frac{k}{A^2}\Bigg{]}\, ,
\end{eqnarray}
where prime denotes the derivative with respect to $z$ and overdot
stands for the derivative with respect to $t$.
 In the bulk, the equations of motion is just the Einstein
equations with a cosmological constant
\begin{equation}
{}^{(n+1)}G_{\mu\nu}=-\kappa_{n+1}^2\Lambda_{n+1}g_{\mu\nu}\, .
\end{equation}
From ${}^{(n+1)}G_{0z}=0$, we obtain
\begin{equation}
\frac{N'}{N}\frac{\dot{A}}{A}=\frac{\dot{A}'}{A}\, .
\end{equation}
Defining
\begin{equation}
F(t,z)=A^{n-2}\left({A'}^2-\frac{\dot{A}^2}{N^2}-k\right)\, ,
\end{equation}
 we have
\begin{equation}
F'=-(n-2)\frac{A'A^{n-1}}{N^2}\left[\frac{\dot{A}^2}{A^2}-N^2\left(\frac{{A'}^2}{A^2}
+\frac{2}{n-2}\frac{A''}{A}\right)+k\frac{N^2}{A^2}\right]\, ,
\end{equation}
\begin{equation}
\dot{F}=(n-2)\dot{A}A^{n-1}\left[\frac{{A'}^2}{A^2}+\frac{2}{n-2}\frac{A'}{A}\frac{N'}{N}
-\frac{1}{N^2}\left(\frac{\dot{A}^2}{A^2}
-\frac{2}{n-2}\frac{\dot{A}}{A}\frac{\dot{N}}{N}
+\frac{2}{n-2}\frac{\ddot{A}}{A}\right)-\frac{k}{A^2}\right]\, .
\end{equation}
In terms of $F$, we can express the $(00)$ and $(zz)$ components
of equations of motion as
\begin{equation}
\label{e66}
 F'=-\frac{2A'A^{n-1}}{n-1}\kappa_{n+1}^2\Lambda_{n+1}\,
, \quad
\dot{F}=\frac{2\dot{A}A^{n-1}}{n-1}\kappa_{n+1}^2\Lambda_{n+1}\, .
\end{equation}
Integrating the first one in (\ref{e66}), we get
\begin{equation}
F+\frac{2\kappa_{n+1}^2\Lambda_{n+1}}{n(n-1)}A^n+\mathcal{C}=0\, ,
\end{equation}
where $\mathcal{C}$ is a function of $t$.  Integrating the second
equation leads to the conclusion that $\mathcal{C}$ is a constant.
Thus we have
\begin{equation}
\label{int1} \left(\frac{\dot{A}}{NA}\right)^2+\frac{k}{A^2}
=\frac{2\kappa_{n+1}^2\Lambda_{n+1}}{n(n-1)}+\left(\frac{A'}{A}\right)^2+\frac{\mathcal{C}}{A^n}\,
.
\end{equation}
We can calculate the Weyl tensor for the metric~(\ref{bulkmetric})
and find that the nonvanishing components of Weyl tensor have the
form
\begin{equation}
C_{\mu\nu\alpha\beta}~\sim~
\frac{1}{N^2}\left(\frac{\dot{A}}{A}\right)^2+\frac{k}{A^2}-\left(\frac{A'}{A}\right)^2
+\frac{A''}{A}-\frac{N''}{N}+\frac{A'}{A}\frac{N'}{N}
+\frac{1}{N^2}\left(\frac{\dot{A}}{A}\frac{\dot{N}}{N}-\frac{\ddot{A}}{A}\right)\,.
\end{equation}
If we impose the constraint
\begin{equation}
\label{constrain}
\frac{A''}{A}=\frac{N''}{N}=\frac{A'}{A}\frac{N'}{N}
+\frac{1}{N^2}\left(\frac{\dot{A}}{A}\frac{\dot{N}}{N}-\frac{\ddot{A}}{A}\right)
=-\frac{2\kappa_{n+1}^2\Lambda_{n+1}}{n(n-1)}=\frac{1}{\ell^2}\, ,
\end{equation}
 we find that the condition with vanishing Weyl tensor exactly
leads to  $\mathcal{C}=0$ in equation~(\ref{int1}). Substituting
this constraint into the Einstein equations, we can see that the
second-order differential equation for the Einstein equations
reduce to a first-order one in the case, which is just the
equation (\ref{int1}) with vanishing $\mathcal{C}$. Therefore to
solve the equation (\ref{int1}) becomes simple when the Weyl
tensor vanishes. Otherwise, one has to solve a second-order
differential equation (Non-linear term will appear in this
second-order differential equation if $n>4$.).

Considering the Israel junction condition on the brane, one can
give the equations of motion on the brane (Friedmann equation)
from (\ref{int1}). The equation ${}^{(n+1)}G_{0z}=0$ also implies
that we have
\begin{equation}
\frac{\dot{A}}{N}=\alpha(t)\, .
\end{equation}
Substituting this equation into the $(00)$ equation
or~(\ref{int1}) with vanishing $\mathcal{C}$, we have
\begin{equation}
\label{eqA}
\alpha^2+k-\left(A'\right)^2-\frac{2\Lambda_{n+1}}{n(n-1)}A^2=0\, .
\end{equation}
Since $\dot{A}(t,0)=\dot{a}(t)$ and $N(t,0)=1$, we have
$\alpha(t)=\dot{a}(t)$. The equation~(\ref{eqA}) has the solution
\begin{equation}
A(t,z)=a(t)\left[-\frac{1}{2}\frac{\ell^2}{\tilde{r}_A^2}
+\left(1+\frac{1}{2}\frac{\ell^2}{\tilde{r}_A^2}\right)\cosh{\left(\frac{2z}{\ell}\right)}
-\sqrt{1+\frac{\ell^2}{\tilde{r}_A^2}}\sinh{\left(\frac{2|z|}{\ell}\right)}\right]^{\frac{1}{2}}\,
.
\end{equation}
One can check that this solution  satisfies the
constraint~(\ref{constrain}). For $n=4$, this solution is just the
one found in~\cite{Langlois} with vanishing $\mathcal{C}$.

The apparent horizon for fixed $z$ has the form
\begin{equation}
\tilde{r}_A|_z=\left(\frac{1}{N^2}\frac{\dot{A}^2}{A^2}+\frac{k}{A^2}\right)^{-\frac{1}{2}}
=\left(\frac{\dot{a}^2}{A^2}+\frac{k}{A^2}\right)^{-\frac{1}{2}}\, ,
\end{equation}
namely, we have
\begin{equation}
\tilde{r}_A|_z=\tilde{r}_A f(z,\tilde{r}_A),
\end{equation}
with $\tilde{r}_A=\tilde{r}_A|_0$ and
\begin{equation}
f(z,\tilde{r}_A)=\left[-\frac{1}{2}\frac{\ell^2}{\tilde{r}_A^2}
+\left(1+\frac{1}{2}\frac{\ell^2}{\tilde{r}_A^2}\right)\cosh{\left(\frac{2z}{\ell}\right)}
-\sqrt{1+\frac{\ell^2}{\tilde{r}_A^2}}\sinh{\left(\frac{2|z|}{\ell}\right)}\right]^{\frac{1}{2}}\,
.
\end{equation}
We  show  this function (the red line for $\ell=1$,
$\tilde{r}_A=0.5$, green one for $\ell=1$, $\tilde{r}_A=2$, and
blue one for $\ell=1$, $\tilde{r}_A=3$ ) in Fig.~1. For any fixed
$\tilde{r}_A$, we find that this function has a zero  at
$z=z_{max}$.  The value $z_{max}$ is determined by solving the
equation $f(z)=0$, which gives
\begin{equation}
z_{max}(\tilde{r}_A)=\ell
\mathrm{arcsinh}\left(\frac{\tilde{r}_A}{\ell}\right)\, .
\end{equation}
This indicates that the apparent horizon extends into the bulk
with a finite distance, $z_{max}$.
\begin{figure}[htb]
\centering
\begin{minipage}[c]{.58\textwidth}
\centering
\includegraphics[width=\textwidth]{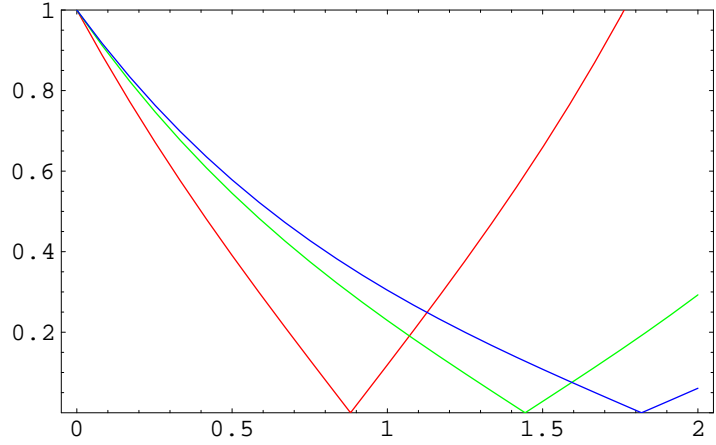}
\caption{The function $f(z)$ has a zero  at $z_{max}$. The red
line is for the case with $\ell=1$, $\tilde{r}_A=0.5$, the green
one for $\ell=1$, $\tilde{r}_A=2$, and the blue one for $\ell=1$,
$\tilde{r}_A=3$, respectively. } \label{ffunction}
\end{minipage}
\end{figure}

Once the bulk geometry is known, we can calculate the area of the
apparent horizon using the bulk geometry. Let us first consider
the case with $n=4$. Considering the $\mathbb{Z}_2$ symmetry, we
have the horizon area
\begin{equation}
\mathcal{A}=2\times 4\pi\tilde{r}_A^2\int _0
^{z_{max}}f^2(z,\tilde{r}_A)dz\, ,
\end{equation}
 Carrying out the integration, we arrive at
\begin{equation}
\mathcal{A}=4\pi \tilde{r}_A^2\ell
\left[\sqrt{1+\frac{\ell^2}{\tilde{r}_A^2}}
-\frac{\ell^2}{\tilde{r}_A^2}\mathrm{arcsinh}\left(\frac{\tilde{r}_A}{\ell}\right)\right]\,
.
\end{equation}
According to the $5$-dimensional area formula, we obtain the entropy
associated with this $5$-dimensional apparent horizon
\begin{equation}
S=\frac{A_2}{4G_5}\ell \left[\sqrt{1+\frac{\ell^2}{\tilde{r}_A^2}}
-\frac{\ell^2}{\tilde{r}_A^2}\mathrm{arcsinh}\left(\frac{\tilde{r}_A}{\ell}\right)\right]\,
.
\end{equation}
Using the relation (\ref{n+1ton}), we can rewrite this entropy as
\begin{equation}
S=\frac{A_2}{4G_4} \left[\sqrt{1+\frac{\ell^2}{\tilde{r}_A^2}}
-\frac{\ell^2}{\tilde{r}_A^2}\mathrm{arcsinh}\left(\frac{\tilde{r}_A}{\ell}\right)\right]\,
.
\end{equation}
This is exactly the form (\ref{e46}) we have given in the previous
section by using the method developed in \cite{caicao}.

For the case with an arbitrary $n$ ($n>4$), the area of the apparent
horizon in the $(n+1)$-dimensional bulk can be expressed as
\begin{equation}
\mathcal{A}=2 \times
\Omega_{n-2}\tilde{r}_A^{n-2}\int_0^{z_{max}(\tilde{r}_A)}f^{n-2}(z,\tilde{r}_A)dz\,
.
\end{equation}
After some calculations, one can obtain
\begin{equation}
\mathcal{A}=\frac{2}{n-1}\Omega_{n-2}\tilde{r}_A^{n-1}
{}_2F_1\left[\frac{n-1}{2},\frac{1}{2},\frac{n+1}{2},-\left(\frac{\tilde{r}_A}{\ell}\right)^2\right]\,
.
\end{equation}
By using the $(n+1)$-dimensional area formula of horizon entropy,
we have
\begin{equation}
S=\frac{\mathcal{A}}{4 G_{n+1}}=\frac{A_{n-2}}{4
G_{n+1}}\frac{2\ell}{n-1}\left(\frac{\tilde{r}_A}{\ell}\right)
{}_2F_1\left[\frac{n-1}{2},\frac{1}{2},\frac{n+1}{2},-\left(\frac{\tilde{r}_A}{\ell}\right)^2\right]\,
.
\end{equation}
Using the relation (\ref{n+1ton}), once again, we arrive at
\begin{equation}
S=\frac{A_{n-2}}{4
G_{n}}\Bigg{\{}\frac{n-2}{n-1}\left(\frac{\tilde{r}_A}{\ell}\right)
{}_2F_1
\left[\frac{n-1}{2},\frac{1}{2},\frac{n+1}{2},-\left(\frac{\tilde{r}_A}{\ell}\right)^2\right]\Bigg
{\}}\, .
\end{equation}
This is completely the same as the entropy expression (\ref{e45})
associated with the apparent horizon on the brane.

Clearly, when the bulk Weyl tensor is nonvanished, one has to
solve a nonlinear second-order differential equation for $n>4$, in
order to give the bulk geometry. Solving this nonlinear equation
is not an easy matter, but it is worth trying. Once the bulk
geometry is given, one can obtain the entropy of apparent horizon
by using the area formula in the bulk.


\section{Conclusion and discussion}

In this paper we have discussed thermodynamics of the apparent
horizon of an $n$-dimensional FRW universe in the RSII brane world
scenario. The gravity theory on the brane  is not the Einstein one
due to the existence of extra dimension, therefore the well-known
area formula for horizon entropy is not applicable on the brane.
By using the method we have developed in~\cite{caicao}, we have
obtained an explicit entropy expression of the apparent horizon in
the $n$-dimensional FRW universe embedded in an
$(n+1)$-dimensional pure AdS space-time. In this model, the
effective energy-momentum tensor coming from the non-Einstein part
is just the $\pi_{\mu\nu}$ in the effective field equations on the
brane. From this effective energy-momentum tensor, we define an
effective energy-supply. By using the unified first law, together
with the Clausius relation $\delta Q=\langle A\stackrel{m}{\Psi}
,\xi\rangle=\frac{\kappa}{8\pi G}\langle dA,\xi\rangle-\langle
A\stackrel{e}{\Psi} ,\xi\rangle=Td S$ (This also suggests that the
associated thermodynamics with the apparent horizon is an
equilibrium one), we have obtained an analytic
expression~(\ref{entropy}) of the entropy for the apparent
horizon. The entropy expression has expected properties. In the
limit of large horizon radius, this entropy reduces to the
 $n$-dimensional area formula, where the Einstein gravity on the brane
 is recovered. In this case, this entropy
can be interpreted as the area of the ``cylinder" as described in
\cite{Emparan}. On the other hand, in the limit of  small horizon
radius, the entropy reduces to $(n+1)$-dimensional area formula,
as respected, and can be interpreted as the area of two ``disks",
which is negligible in the large radius limit. In addition, we
have shown that the Friedmann equation on the brane can be
rewritten as a universal form like the first law, $dE=TdS +W dV$.

Our entropy expression for the apparent horizon in FRW universe is
useful in the study of thermodynamics of black holes on the brane
since we expect that the entropy associated with apparent horizon
and black hole horizon has a same expression.

We have also discussed the bulk geometry for an arbitrary
dimension $n\geq 4$, and given a solution with vanishing Weyl
tensor.  We have found that the apparent horizon extends into the
bulk with a finite distance which are labelled by
$z_{max}(\tilde{r}_A)=\ell
\mathrm{arcsinh}\left(\tilde{r}_A/\ell\right)$.  Using the bulk
geometry and the area formula in the bulk, we  have calculated the
entropy associated with the apparent horizon extended into the
bulk. We have found that both methods give the completely same
expression for the apparent horizon entropy.  It is of great
interest to extend our method to other brane world scenarios such
as DGP model, models with a bulk Gauss-Bonnet term, etc.

\section*{Acknowledgments}
This research was initiated during R.G.Cai's visit to the department
of physics, Fudan university, the warm hospitality extended to him
is appreciated. R.G. Cai thanks B. Wang and R.K. Su for helpful
discussions.  This work was supported partially by grants from NSFC,
China (No. 10325525 and No. 90403029), and a grant from the Chinese
Academy of Sciences.

\end{document}